\documentclass{ws-ijmpcs}

\def\trimmarks{}

\usepackage[english]{babel}
\usepackage{enumerate}
\usepackage{amsmath,amssymb}
\usepackage{letltxmacro}			
\usepackage{xkeyval}				
\usepackage{xargs}					
\usepackage{xifthen}				
\usepackage{slashed}
\usepackage[normalem]{ulem}
\usepackage{bm}					
\usepackage{dsfont}
\usepackage[super,sort&compress]{natbib}
\usepackage[cmyk,usenames,dvipsnames,svgnames,table,x11names]{xcolor}
\usepackage{tikz}
\usepackage{tikz-3dplot}
\usetikzlibrary{calc,matrix, shapes,arrows,decorations.pathmorphing,decorations.markings,decorations.pathreplacing,trees}
\usepackage[pdftex, unicode=true, pdfnewwindow=true, colorlinks=true, linkcolor=black, urlcolor=black, citecolor=black, hyperindex, pdfauthor={Frederik F. Van der Veken}, pdftitle={Working with Piecewise Linear Wilson Lines}, bookmarksopen=true, bookmarksopenlevel=0]{hyperref}

\makeatletter
	\let\oldr@@t\r@@t
	\def\r@@t#1#2{%
	\setbox0=\hbox{$\oldr@@t#1{#2\,}$}\dimen0=\ht0
	\advance\dimen0-0.2\ht0
	\setbox2=\hbox{\vrule height\ht0 depth -\dimen0}%
	{\box0\lower0.4pt\box2}}
	\LetLtxMacro{\oldsqrt}{\sqrt}
	\renewcommand*{\sqrt}[2][\ ]{\oldsqrt[#1]{#2}}
\makeatother

\newcommandx*{\1}[2][1,2]{\ensuremath{
		\ifthenelse{\isempty{#1} \AND \isempty{#1}}{\mathds{1}}{
			\ifthenelse{\isempty{#2}}
				{\mathds{1}_{#1}}
				{\mathds{1}_{{#1}\times {#2}}}
}
}}

\newcommand{\ep}{\, .}
\newcommand{\ec}{\, ,}

\renewcommand*{\bar}{\overline}

\newcommand{\Not}{\stackrel{\text{\tiny N}}{=}}							
\renewcommand{\exp}[1]{\ensuremath{\text{e}^{#1}}}

\renewcommand*{\i}{\text{i}\hspace*{1pt}}

\newlength{\nplength}
\newcommandx{\negphantom}[1]{\settowidth{\nplength}{#1}\hspace{-\nplength}}

\renewcommandx*{\vec}[3][2, 3]{{\ensuremath{\mathbf{\bm{#1}}_\mathrm{#2}^{\mathrm{#3}}}}}
\newcommandx*{\Trace}[1][1]{\ensuremath{
		\:\textrm{Tr}\ifthenelse{\isempty{#1}}{\,}{\!\left( #1 \right)}
}}
\newcommandx*{\trace}[1][1]{\ensuremath{
		\:\textrm{tr}\ifthenelse{\isempty{#1}}{\,}{\!\left( #1 \right)}
}}

\newcommandx*{\diracdelta}[2][1]{\ensuremath{
		\ifthenelse{\isempty{#1}}
		{\,\delta\!\left({#2}\right)}
		{\,\delta^{\IfInteger{#1}{({#1})}{{#1}}}\!\left({#2}\right)}
}}
\newcommand{\heavisidetheta}[1]{\, \theta\!\left( #1 \right)}

\newcommandx*{\Int}[4][1, 2, 4={0pt}, usedefault, addprefix=\global]{
	\ensuremath{\int\limits_{\:\!{#1}}^{\:\!{#2}}{\!\!}\hspace{#4} {\protect#3}\,\,}
}

\newcommandx*{\Dif}[2][1]{\ensuremath{{\textrm{d}}^{{#1}} {#2}\,}}
\newcommandx*{\dif}[2][1]{\ensuremath{\partial^{{#1}} {#2}\,}}
\newcommandx*{\MDif}[2][1]{
	\ensuremath{\frac{{\textrm{d}}^{{#1}} {#2}\,}{
		\ifthenelse{\isempty{#1}}{2\pi} { \left(2\pi\right)^{{#1}} }
	}}
}
\newcommandx*{\DDif}[2][1]{\ensuremath{{\mathcal{D}}^{{#1}} {#2}\,}}
\newcommandx*{\Diff}[8][1, 4, 5, 6, 7, 8]{\ensuremath{
		\ifthenelse{\isempty{#4} \AND \isempty{#5}}
			{\frac{\Dif[#1]{#2}}{{\Dif{#3}}^{{#1}}}}
			{\frac{\Dif[#1]{#2}}{
				\ifthenelse{\isempty{#4}} {\Dif{#3}} {\left(\Dif{#3}\right)^{#4}}
				\ifthenelse{\isempty{#5}} {} { \ifthenelse{\isempty{#6}} {\Dif{#5}} {\left(\Dif{#5}\right)^{#6}}}
				\ifthenelse{\isempty{#7}} {} { \ifthenelse{\isempty{#8}}{\Dif{#7}}{\left(\Dif{#7}\right)^{#8}}}
}}}}
\newcommandx*{\diff}[8][1, 4, 5, 6, 7, 8]{\ensuremath{
		\ifthenelse{\isempty{#4} \AND \isempty{#5}}
			{\frac{\dif[#1]{#2}}{{\dif{#3}}^{{#1}}}}
			{\frac{\dif[#1]{#2}}{
				\ifthenelse{\isempty{#4}} {\dif{#3}} {\left(\dif{#3}\right)^{#4}}
				\ifthenelse{\isempty{#5}} {} { \ifthenelse{\isempty{#6}} {\dif{#5}} {\left(\dif{#5}\right)^{#6}}}
				\ifthenelse{\isempty{#7}} {} { \ifthenelse{\isempty{#8}}{\dif{#7}}{\left(\dif{#7}\right)^{#8}}}
}}}}

\newcommandx*{\braket}[3][1,2,3]{
	\ifthenelse{\isempty{#2} \AND \isempty{#3}}
	{\ifthenelse{\isempty{#1}}  {\opm{!! empty braket used !!}}  {\ensuremath{\left< {#1} \right>} } }
	{\ifthenelse{\isempty{#3}}
			{\ensuremath{\left< {#1} \vphantom{{#2}}  \:\! \right| \! \! \! \; \left. {#2} \vphantom{{#1}} \right>} }
			{\ensuremath{\left< {#1} \vphantom{{#3}} \right| #2 \left| {#3} \vphantom{{#1}} \right>} }
}}

\newcommandx*{\wilson}[3][3=]{\ensuremath{
	\,\ifthenelse{\isempty{#1} \AND \isempty{#2}}	{\mathcal{U}^{#3}}	{\mathcal{U}_{(#1 \, ; \,  #2)}^{#3}}
}}

\newcommand{\wilsonup}{
	\def\myscale{0.6}
	\begin{tikzpicture}[scale=0.6, baseline= {($(current bounding box.base)-(0,1.7pt)$)}]
		\draw[wilson,-implies] (0,0) -- (1.5,0);
		\filldraw[wilsontext] (0,0) circle(0.125);
	\end{tikzpicture}
}
\newcommand{\wilsondown}{
	\def\myscale{0.6}
	\begin{tikzpicture}[scale=0.6, baseline= {($(current bounding box.base)-(0,1.7pt)$)}]
		\draw[wilson,implies-] (0,0) -- (1.5,0);
		\filldraw[wilsontext] (1.5,0) circle(0.125);
	\end{tikzpicture}
}
\newcommand{\wilsonupreversed}{
	\def\myscale{0.6}
	\begin{tikzpicture}[scale=0.6, baseline= {($(current bounding box.base)-(0,1.7pt)$)}]
		\draw[wilson] (0,0) -- (1.5,0);
		\draw[wilson,-implies] (1.56,0) -- (1.4,0);
		\filldraw[wilsontext] (0,0) circle(0.125);
	\end{tikzpicture}
}
\newcommand{\wilsondownreversed}{
	\def\myscale{0.6}
	\begin{tikzpicture}[scale=0.6, baseline= {($(current bounding box.base)-(0,1.7pt)$)}]
		\draw[wilson] (0,0) -- (1.5,0);
		\draw[wilson,-implies] (-0.06,0) -- (0.1,0);
		\filldraw[wilsontext] (1.5,0) circle(0.125);
	\end{tikzpicture}
}

\newenvironment{subalign}[1][]{
	\subequations
	\ifthenelse{\isempty{#1}}{}{\label{#1}}
	\align
}{
	\endalign
	\endsubequations
}

\definecolor{geel}{cmyk}{0.,0.075,1.,0.2}
\definecolor{blauw}{cmyk}{1.,0.55,0.05,0.05}
\definecolor{groen}{cmyk}{0.6,0.15,1.,0.1}
\definecolor{rood}{cmyk}{0.1,1.,0.1,0.3}
\newcommand{\photoncolour}{geel}
\newcommand{\gluoncolour}{groen}
\newcommand{\fermioncolour}{geel}
\newcommand{\darkfermioncolour}{geel!90!blauw!80!black}
\newcommand{\wilsoncolour}{blauw}
\newcommand{\eikonalcolour}{rood}
\newcommand{\bloboutercolour}{gray!80!black}
\newcommand{\blobinnercolour}{gray!20}
\newcommand{\pdfoutercolour}{geel!80!groen!80!blauw}
\newcommand{\pdfinnercolour}{geel!60!groen!80!blauw!20!white}

\pgfdeclaredecoration{complete sines}{initial}{				
	\state{initial}[
		width=+0pt,
		next state=sine,
		persistent precomputation={
			\pgfmathsetmacro\matchinglength{
				\pgfdecoratedinputsegmentlength / int(\pgfdecoratedinputsegmentlength/\pgfdecorationsegmentlength)
			}
			\setlength{\pgfdecorationsegmentlength}{\matchinglength pt}
        		}
	] {}
	\state{sine}[width=\pgfdecorationsegmentlength]{
		\pgfpathsine{\pgfpoint{0.25\pgfdecorationsegmentlength}{0.5\pgfdecorationsegmentamplitude}}
		\pgfpathcosine{\pgfpoint{0.25\pgfdecorationsegmentlength}{-0.5\pgfdecorationsegmentamplitude}}
		\pgfpathsine{\pgfpoint{0.25\pgfdecorationsegmentlength}{-0.5\pgfdecorationsegmentamplitude}}
		\pgfpathcosine{\pgfpoint{0.25\pgfdecorationsegmentlength}{0.5\pgfdecorationsegmentamplitude}}
	}
	\state{final}{}
}
\pgfdeclaredecoration{integralshape}{integralshape}{		
	\state{integralshape}[width=+\pgfdecoratedremainingdistance,next state=final] {
		\pgfpathmoveto{\pgfpoint{0}{0.042*\pgfdecoratedpathlength*\pgfdecorationsegmentamplitude}}
		\pgfpathcurveto
			{\pgfpoint{-0.01*\pgfdecoratedpathlength*\pgfdecorationsegmentamplitude}{0.02*\pgfdecoratedpathlength*\pgfdecorationsegmentamplitude}}
			{\pgfpoint{\pgfdecoratedpathlength*(0.5-0.18*\pgfdecorationsegmentamplitude)}{0.18*tan(\pgfdecorationsegmentangle)*\pgfdecorationsegmentamplitude*\pgfdecoratedpathlength}}
			{\pgfpoint{\pgfdecoratedpathlength*0.5}{0}}
		\pgfpathcurveto
			{\pgfpoint{\pgfdecoratedpathlength*(0.5+0.18*\pgfdecorationsegmentamplitude)}{-0.18*tan(\pgfdecorationsegmentangle)*\pgfdecorationsegmentamplitude*\pgfdecoratedpathlength}}
			{\pgfpoint{\pgfdecoratedpathlength*(1+0.01*\pgfdecorationsegmentamplitude)}{-0.02*\pgfdecoratedpathlength*\pgfdecorationsegmentamplitude}}
			{\pgfpoint{\pgfdecoratedpathlength}{-0.042*\pgfdecoratedpathlength*\pgfdecorationsegmentamplitude}}
	}
	\state{final}{}
}

\def\myscale{1}
\tikzset{
photon/.style={decorate, decoration={snake,amplitude=\myscale*3pt, segment length=\myscale*7pt}, draw=\photoncolour, line width=\myscale*1.4pt},
gluon/.style={decorate, draw=\gluoncolour, decoration={coil,amplitude=\myscale*3pt, segment length=\myscale*4pt},  line width=\myscale*1.2pt},
quark/.style={draw=\fermioncolour, postaction={decorate},
	decoration={markings,mark=at position .5*\pgfdecoratedpathlength+sqrt(\myscale)*5.5pt with {\arrow[\fermioncolour]{latex}}},  line width=\myscale*1.6pt},		
quarknoarrow/.style={draw=\fermioncolour, line width=\myscale*1.6pt},
darkquark/.style={draw=\darkfermioncolour, postaction={decorate},
	decoration={markings,mark=at position .5*\pgfdecoratedpathlength+sqrt(\myscale)*5.5pt with {\arrow[\darkfermioncolour]{latex}}},  line width=\myscale*1.6pt},
eikonal/.style={double, double distance=\myscale*2.5pt, draw=\eikonalcolour, postaction={decorate},
	decoration={markings,mark=at position .5*\pgfdecoratedpathlength+sqrt(\myscale)*7.5pt with {\arrow[\eikonalcolour]{angle 60}}}, line width=\myscale*1.2pt},
wilson/.style={double, double distance=\myscale*2.5pt, line width=\myscale*1.2pt, draw=\wilsoncolour},
blob/.style={draw=\bloboutercolour, fill=\blobinnercolour, line width=\myscale*1.2pt},
pdf/.style={draw=\pdfoutercolour, fill=\pdfinnercolour, line width=\myscale*1.2pt},
photontext/.style={\photoncolour!80!black},
gluontext/.style={\gluoncolour!80!black},
quarktext/.style={\fermioncolour!80!black},
wilsontext/.style={\wilsoncolour!80!black},
eikonaltext/.style={\eikonalcolour!80!black},
blobtext/.style={\bloboutercolour!80!black},
pdftext/.style={\pdfoutercolour!80!black},
accolade/.style={gray,decorate, decoration={brace,amplitude=\myscale*5pt}, line width=\myscale*1.2pt},
hide/.style={ultra thick, white},
finalstatecut/.style={decorate, decoration={integralshape,amplitude=(1/\myscale)*1.5pt,angle=2}, line width=\myscale*1.2pt},
wilsonarrow/.style={postaction={decorate}, decoration={markings,mark=at position .75 with {\arrow[\wilsoncolour]{angle 60}}}},
wilsonarrowreversed/.style={postaction={decorate}, decoration={markings,mark=at position .3 with {\arrowreversed[\wilsoncolour]{angle 60}}}},
wilsonarrow2/.style={postaction={decorate}, decoration={markings,mark=at position 1. with {\arrow[\wilsoncolour]{angle 60}}}},
wilsonarrowreversed2/.style={postaction={decorate}, decoration={markings,mark=at position 0 with {\arrowreversed[\wilsoncolour]{angle 60}}}}
}
\newenvironment{tikzfigure}[2]{
	\def\myscale{#1}
	\begin{tikzpicture}[baseline= {($(current bounding box.base)-(0pt,#2)$)},scale=\myscale]
}
{
	\end{tikzpicture}
}

\begin{document}

\markboth{Frederik~F.~Van~der~Veken}
{Piecewise linear Wilson lines}

%
\catchline{}{}{}{}{}
%

\title{A new approach to piecewise linear Wilson lines}

\author{Frederik~F.~Van~der~Veken}

\address{Department of Physics, University of Antwerp, Groenenborgerlaan 171, 2020 Antwerp, Belgium\\
frederikvanderveken@gmail.com}

\maketitle


\begin{abstract}
Wilson lines are key objects in many QCD calculations. They are parallel transporters of the gauge field that can be used to render non-local operator products gauge invariant, which is especially useful for calculations concerning validation of factorization schemes and in calculations for constructing or modelling parton density functions. We develop an algorithm to express Wilson lines that are defined on piecewise linear paths in function of their Wilson segments, reducing the number of diagrams needed to be calculated. We show how different linear path topologies can be related using their color structure. This framework allows one to easily switch results between different Wilson line structures, which is helpful when testing different structures against each other, e.g.\@ when checking universality properties of non-perturbative objects.
\end{abstract}

\ccode{PACS numbers: 11.15.Tk, 12.38.Aw, 12.38.Lg.}

\section{Introduction}
A general Wilson line is an exponential of gauge fields along a path $\mathcal{C}$, defined as:
\begin{equation}\label{eq: Wilsondef}
  \mathcal{U} =
    \mathcal{P}\,\exp{\i g \int_\mathcal{C} \Dif{z^\mu} A_\mu (z)}\ep
\end{equation}
Because the gauge fields are non-Abelian, i.e.\@ $A_\mu = t^a A^a_\mu$ where $t^a$ is a generator of a Lie algebra, they have to be \emph{path ordered}, denoted by the symbol $\mathcal{P}$ in \eqref{eq: Wilsondef}. The fields are ordered such that the fields first on the path are written leftmost. After making a Fourier transform, the path content is fully described by the following integrals:
\begin{equation} \label{eq: PathIntegrals}
  I_n =
    \frac{1}{n!}\; 
    \mathcal{P} \!\!\int\! \Dif{\lambda_1}\cdots \Dif{\lambda_n}
    \left(z_1^{\mu_1}\right)' \cdots \left(z_n^{\mu_n}\right)'
    \exp{\i\prod\limits^n k_i\cdot z_i}\ec
\end{equation}
so that the $n$-th order term of the Wilson line expansion is given by
\begin{equation}
  \mathcal{U}_n =
    \left(\i g\right)^n \! \Int{\MDif[\omega]{k_1}\cdots\MDif[\omega]{k_n}}
      A_{\mu_n}(-k_n)\!\cdots\! A_{\mu_1}(-k_1) \; I_n\ep
\end{equation}
An important property is its gauge transform, only dependent on its endpoints $a^\mu$ and $b^\mu$:
\begin{equation}\label{eq: gaugetransform}
  \wilson{b}{a} \rightarrow
    \exp{\i g\alpha(b)} \wilson{b}{a} \, \exp{-\i g \alpha(a)} \ep
\end{equation}
This property is used to give a gauge invariant operator definition for transverse momentum dependent parton density functions. As the gauge transformation of the Wilson line only depends on its endpoints, there is some freedom on the choice of the path. The correct path can then be constructed by identifying the appropriate gluon radiation for the given process.\cite{Collins:1981uk,Collins:1982wa,Collins:1984kg,Belitsky:2002sm,Boer:2003cm,Ji:2004xq,Bacchetta:2004jz,Belitsky:2005qn,Hautmann:2007uw,Cherednikov:2009wk,Cherednikov:2011ku,Collins:2011zzd,Boer:2011fh,MertAybat:2011wy,Echevarria:2012js,Bacchetta:2012qz,Collins:2013aa}

In the so-called \emph{eikonal approximation},  a moving quark is considered to emit only soft and collinear radiation, which can be resummed into a Wilson line. One case where we could use the eikonal approximation, is for a quark in a dense medium (see e.g.\@ Refs.~[\refcite{Blaizot:2012fh}] and [\refcite{Iancu:2013dta}] where in the latter the medium is reduced to a shockwave). Other applications of Wilson lines include the calculation of soft factors \cite{GarciaEchevarria:2011rb,Becher:2012za,Idilbi:2007ff,Chiu:2009yx}, the study of jet quenching \cite{Cherednikov:2013pba}, a recast of QCD in loop space where the geometric evolution of rectangular loops can be related to its energy evolution \cite{Mertens:2014hma,Cherednikov:2012qq,VanderVeken:2014lka,Mertens:2014lla,Mertens:2014mia}, etc.

\section{Linear Path Segments}
We start our calculations by quickly reviewing the calculation of one linear segment, which we will extend in Sec.\@ \ref{sec: PiecewiseLinear} to a set of linear segments. For every segment there exist four possible path structures: it can be a finite segment connecting two points, it can be a semi-infinite line connecting $\pm\infty$ to a point $r^\mu$, or it can be a fully infinite line connecting $-\infty$ and $+\infty$. In this paper we won't consider the last case.
We start by considering a line from a point $a^\mu$ to $+\infty$, along a direction $\hat{n}^\mu$:
\begin{equation}
  z^\mu = a^\mu + \lambda \, \hat{n}^\mu \quad \lambda = 0 \ldots \infty\ep
\end{equation}
Using \eqref{eq: PathIntegrals}, it is straightforward to calculate the segment integrals:
\begin{equation}
  I_n^\text{l.b.} =  \label{eq: lowerbound}
    \hat{n}^{\mu_1}\cdots \hat{n}^{\mu_n} \,
    \exp{\i a\cdot \sum\limits_j k_j} \,
    \prod\limits_{j=1}^n
    \frac{\i}{\hat{n}\!\cdot\! \sum\limits_{l=j}^n k_l + \i \eta}\ep
\end{equation}
The $\i\eta$ are convergence terms added to regularize the exponent.
%
\begin{figure*}[t!]
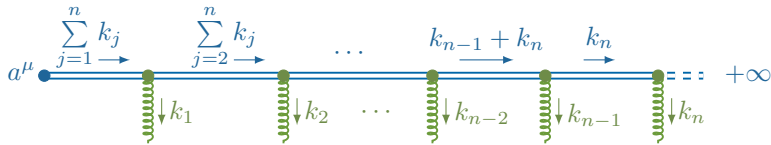

\centering
\begin{tikzfigure}{0.6}{0}
  \draw[wilson] (1.8,0) -- (15.3,0);
  \draw[wilson,dashed] (15.5,0) -- (16.3,0);
  \node[wilsontext] at (17.3,0.03) {$+\infty$};
  \filldraw[wilsontext] (1.7,0) circle(0.125);
  \node[wilsontext] at (1.2,0.1) {$a^\mu$};
  
  \draw[gluon] (4,0) -- (4,-1.5);
  \filldraw[gluontext] (4,0) circle (0.125);
  \draw[-latex, gluontext] (4.3,-0.5) -- (4.3, -1.1);
  \node[gluontext] at (4.75,-0.8) {$k_1$};
  \draw[-latex,wilsontext] (2.9,0.35) -- (3.6,0.35);
  \node[wilsontext] at (2.7,0.85) {$\sum\limits_{j=1}^{n} k_j $};
  \begin{scope}[shift={(3,0)}]
    \draw[gluon] (4,0) -- (4,-1.5);
    \filldraw[gluontext] (4,0) circle (0.125);
    \draw[-latex, gluontext] (4.3,-0.5) -- (4.3, -1.1);
    \node[gluontext] at (4.75,-0.8) {$k_2$};
    \draw[-latex,wilsontext] (2.9,0.35) -- (3.6,0.35);
    \node[wilsontext] at (2.7,0.85) {$\sum\limits_{j=2}^{n} k_j $};
  \end{scope}
  \node[wilsontext] at (8.5,0.5) {$\cdots$};
  \node[gluontext] at (9.1,-0.8) {$\cdots$};
  \begin{scope}[shift={(6.3,0)}]
    \draw[gluon] (4,0) -- (4,-1.5);
    \filldraw[gluontext] (4,0) circle (0.125);
    \draw[-latex, gluontext] (4.3,-0.5) -- (4.3, -1.1);
    \node[gluontext] at (5.1,-0.85) {$k_{n-2}$};
  \end{scope}
  \begin{scope}[shift={(8.8,0)}]
    \draw[gluon] (4,0) -- (4,-1.5);
    \filldraw[gluontext] (4,0) circle (0.125);
    \draw[-latex, gluontext] (4.3,-0.5) -- (4.3, -1.1);
    \node[gluontext] at (5.1,-0.9) {$k_{n-1}$};
    \draw[-latex,wilsontext] (2.1,0.35) -- (3.3,0.35);
    \node[wilsontext] at (2.7,0.8) {$ k_{n-1}+k_n $};
  \end{scope}
  \begin{scope}[shift={(11.3,0)}]
    \draw[gluon] (4,0) -- (4,-1.5);
    \filldraw[gluontext] (4,0) circle (0.125);
    \draw[-latex, gluontext] (4.3,-0.5) -- (4.3, -1.1);
    \node[gluontext] at (4.75,-0.8) {$k_n$};
    \draw[-latex,wilsontext] (2.35,0.35) -- (3.05,0.35);
    \node[wilsontext] at (2.7,0.85) {$k_n $};
  \end{scope}
\end{tikzfigure}
\caption{$n$-gluon radiation for a Wilson line going from $a^\mu$ to $+\infty$.}
\label{fig: lowerbound}
\end{figure*}
As an illustration, the resulting $n$-th order diagram is drawn in Fig.\@ \ref{fig: lowerbound}.

Next we investigate a path that starts at $-\infty$ and goes up to a point $b_\mu$:
\begin{equation}
  z^\mu = b^\mu  + \hat{n}^\mu \, \lambda \qquad \lambda = -\infty \ldots 0\ep
\end{equation}
The resulting path integral is almost the same as before:
\begin{equation}\label{eq: upperbound}
  I_n^\text{u.b.} =
    \hat{n}^{\mu_1}\cdots \hat{n}^{\mu_n} \,
    \exp{\i b\cdot \sum\limits_j k_j} \,
    \prod\limits_{j=1}^n
    \frac{-\i}{\hat{n}\!\cdot\! \sum\limits_{l=1}^j k_l - \i \eta} \ec
\end{equation}
which differs from \eqref{eq: lowerbound} only in the accumulation of momenta in the denominators (and the sign of the convergence terms).
Also note that reversing the path flow is the same as taking the Hermitian conjugate.\cite{VanderVeken:2014kna}

We now introduce a shorthand notation to denote the path structure for a Wilson line segment. We represent the two structures we already calculated as:
\begin{subalign}
  \wilson{+\infty}{a} &\quad \Not \quad \wilsonup \ec \\
  \wilson{b}{-\infty} &\quad \Not \quad \wilsondownreversed \ep 
\end{subalign}
Path reversing equals changing the type and flipping $\hat{n}$.\cite{VanderVeken:2014kna} We represent this as:
\begin{subalign}
  \wilson{a}{+\infty} = \wilson{a}{-\infty}\Big|_{\hat{n}\,\rightarrow\,-\hat{n}} &\quad \Not \quad  \wilsondown \ec \\
  \wilson{-\infty}{b} = \wilson{+\infty}{b}\Big|_{\hat{n}\,\rightarrow\,-\hat{n}} &\quad \Not \quad \wilsonupreversed \ep 
\end{subalign}
A nice feature of this notation is that we get a symbolic ``mirror relation'':
\begin{equation*}
  \left( \wilsonup\right)^\dagger
  = \,
  \wilsondown
  \ec
  \qquad
  \left(\wilsondownreversed\right)^\dagger
  = \,
  \wilsonupreversed
  \ep
\end{equation*}
Next we investigate a Wilson line on a finite path:
\begin{equation}
  z^\mu =
    a^\mu  + \hat{n}^\mu \lambda \qquad \lambda = 0 \, \ldots \, \left\lVert b-a \right\rVert.
\end{equation}
Dropping the factor in front of the integral, we find a recursion relation:
\begin{equation}\label{eq: recursionrelation}
  I_n^\text{fin} {\scriptstyle\left(k_1,\ldots,k_n\right) } =
    \frac{\i}{\hat{n} \!\cdot\! k_1 } \!\! \left(\vphantom{\frac{\i}{k_n}} \!
      I_{n-1}^\text{fin.} {\scriptstyle\left(k_1 + k_2,\ldots,k_n\right)} -
      I_{n-1}^\text{fin.} {\scriptstyle\left(k_2,\ldots,k_n\right)} \!
    \right)
\end{equation}
which we can solve exactly by careful inspection (reintroducing the factor in front):
\begin{equation}\label{eq: finiteResult}
  I_n^\text{fin.} =
    \hat{n}^{\mu_1}\cdots \hat{n}^{\mu_n} \,
    \sum_{m=0}^n
    \left(
	    \prod\limits_{j=1}^m \frac{\i\, \exp{\i a\cdot k_j}}
	    	{\hat{n} \!\cdot\!\sum\limits_{l=j}^m k_l }
	\right)\left(
    	\prod\limits_{j=m+1}^n \frac{-\i\, \exp{\i b\cdot k_j}}
    		{\hat{n}\!\cdot\! \!\!\!\!\sum\limits_{l=m+1}^{j}k_l }
    \right) \ep
\end{equation}
This kind of chained sum can in general be rewritten as a product of two infinite sums. 
Hence we can transform equation \eqref{eq: finiteResult} into a product of two semi-infinite lines:
\begin{equation}
  \wilson{b}{a} = \wilson{+\infty}{b}[\dagger]\wilson{+\infty}{a} = \wilson{b}{-\infty}\wilson{a}{-\infty}[\dagger],
\end{equation}
which can be illustrated schematically as:
\begin{equation}
  \begin{tikzfigure}{0.6}{0.08}
    \draw[wilson, wilsonarrow] (0,0) -- (2,0);
    \filldraw[wilsontext] (0,0) circle(0.125);
    \filldraw[wilsontext] (2,0) circle(0.125);
  \end{tikzfigure}
  \quad = \quad \wilsonup \,\otimes \, \wilsondown
  \quad = \quad \wilsonup \, \otimes \left(\wilsonup\right)^\dagger \ep
\end{equation}

\section{Relating  Different Path Topologies}
We can relate all six path structures to each other. If we choose
\begin{equation}\label{eq: basicstructures}
  \wilsonup\ec \, \wilsondown
\end{equation}
as basic structures, we can express the remaining four as:
\begin{align*}
  \wilsondownreversed &\; = \;
  \wilsondown\big|_{\hat{n}\rightarrow - \hat{n}}
  \ec &
  \wilsonupreversed &\; = \;
  \wilsonup\big|_{\hat{n}\rightarrow - \hat{n}}
  \ec \\
  \begin{tikzfigure}{0.6}{2pt}
    \draw[wilson,wilsonarrow] (0,0) -- (2,0);
    \filldraw[wilsontext] (0,0) circle(0.125);
    \filldraw[wilsontext] (2,0) circle(0.125);
    \node[left] at (0,0.075) {$a^\mu$};
    \node[right] at (2,0.075) {$b^\mu$};
  \end{tikzfigure} &\!\!\!\:=
  a^\mu\wilsonup \otimes \wilsondown\, b^\mu\ec &
  \begin{tikzfigure}{0.6}{2pt}
    \draw[wilson,wilsonarrowreversed] (0,0) -- (2,0);
    \filldraw[wilsontext] (0,0) circle(0.125);
    \filldraw[wilsontext] (2,0) circle(0.125);
    \node[left] at (0,0.075) {$a^\mu$};
    \node[right] at (2,0.075) {$b^\mu$};
  \end{tikzfigure} &\!\!\!\:=
  b^\mu\wilsonup \otimes \wilsondown \,a^\mu\ep
\end{align*}
The two basic structures aren't fully independent either, as they are related by a sign difference and an interchange of momentum indices:
\begin{equation}
  \wilsondown \; = \; (-)^n \,
  \wilsonup
  \big|_{\left(k_1,\ldots,k_n\right) \rightarrow \left(k_n,\ldots,k_1\right)}\ep
\end{equation}
We can use this relation when connecting a Wilson line to a blob, e.g.\@
\begin{equation*}
  \begin{tikzfigure}{0.6}{0.5}
    \draw[gluon] (0.5,0) -- (0.5,-1.5);
    \draw[gluon] (1,0) -- (1,-1.5);
    \draw[gluon] (2.5,0) -- (2.5,-1.5);
    \node[gluontext] at (1.75,-0.45) {$\ldots$};
    \draw[wilson, -implies] (0,0) -- (3,0);
    \filldraw[wilsontext] (0,0) circle(0.125);
    \filldraw[blob] (1.5,-1.5) circle(1.5 and 0.75);
    \node[blobtext] at (1.5,-1.5) {$F$};
  \end{tikzfigure}
  \! = \;
    \left(\i g\right)^n t^{a_n}\!\cdots t^{a_1}\! \Int{\mathcal{D}k}
    I_n^\text{l.b.} \, F_{\mu_1\cdots\mu_n}^{a_1\cdots a_n}{ (k_1,\ldots, k_n)}
    \ep
\end{equation*}
This blob can be constructed from any combination of Feynman diagrams, but cannot contain other Wilson lines. We absorb the gluon propagators into the blob $F^{a_1 \cdots a_n}_{\mu_1\cdots \mu_n}$, and always define the blob as the sum of all possible crossings; it is thus symmetric under the simultaneous interchange of Lorentz, color, and momentum indices. Because every Lorentz index of $F$ is contracted with the same vector $\hat{n}^\mu$, it is automatically symmetric in these. This leads to the fact that an interchange of momentum variables is equivalent to an interchange of the corresponding color indices. It makes it straightforward to relate the two basic structures:
\begin{equation}
  \begin{tikzfigure}{0.6}{0.5}
    \draw[gluon] (0.5,0) -- (0.5,-1.5);
    \draw[gluon] (1,0) -- (1,-1.5);
    \draw[gluon] (2.5,0) -- (2.5,-1.5);
    \node[gluontext] at (1.75,-0.45) {$\ldots$};
    \draw[wilson, implies-] (0,0) -- (3,0);
    \filldraw[wilsontext] (3,0) circle(0.125);
    \filldraw[blob] (1.5,-1.5) circle(1.5 and 0.75);
    \node[blobtext] at (1.5,-1.5) {$F$};
  \end{tikzfigure}
  = (-)^n 
  \begin{tikzfigure}{0.6}{0.5}
    \draw[gluon] (0.5,0) -- (0.5,-1.5);
    \draw[gluon] (1,0) -- (1,-1.5);
    \draw[gluon] (2.5,0) -- (2.5,-1.5);
    \node[gluontext] at (1.75,-0.45) {$\ldots$};
    \draw[wilson, -implies] (0,0) -- (3,0);
    \filldraw[wilsontext] (0,0) circle(0.125);
    \filldraw[blob] (1.5,-1.5) circle(1.5 and 0.75);
    \node[blobtext] at (1.5,-1.5) {$F$};
  \end{tikzfigure}
  \Bigg|_{\left(a_1,\ldots,a_n\right) \rightarrow \left(a_n,\ldots,a_1\right)} \ep
\end{equation}
Often the blob has a factorable color structure, i.e.\@
\begin{equation}
  F_{\mu_1\cdots\mu_n}^{a_1\cdots a_n} (k_1,\ldots, k_n) =
    C^{a_1\cdots a_n} F_{\mu_1\cdots\mu_n} (k_1,\ldots, k_n)\ep
\end{equation}
If we then define the following notations:
\begin{align}
  C &= t^{a_n}\cdots t^{a_1} C^{a_1\cdots a_n}\ec &
  \bar{C} &= t^{a_n}\cdots t^{a_1} C^{a_n\cdots a_1}\ec
\end{align}
we can simply write
\begin{equation}
  \begin{tikzfigure}{0.6}{0.5}
    \draw[gluon] (0.5,0) -- (0.5,-1.5);
    \draw[gluon] (1,0) -- (1,-1.5);
    \draw[gluon] (2.5,0) -- (2.5,-1.5);
    \node[gluontext] at (1.75,-0.45) {$\ldots$};
    \draw[wilson, -implies] (0,0) -- (3,0);
    \filldraw[wilsontext] (0,0) circle(0.125);
    \filldraw[blob] (1.5,-1.5) circle(1.5 and 0.75);
    \node[blobtext] at (1.5,-1.5) {$F$};
  \end{tikzfigure}
  \!=\; C
  \begin{tikzfigure}{0.6}{0.5}
    \draw[photon] (0.5,0) -- (0.5,-1.5);
    \draw[photon] (1,0) -- (1,-1.5);
    \draw[photon] (2.5,0) -- (2.5,-1.5);
    \node[photontext] at (1.75,-0.45) {$\ldots$};
    \draw[wilson, -implies] (0,0) -- (3,0);
    \filldraw[wilsontext] (0,0) circle(0.125);
    \filldraw[blob] (1.5,-1.5) circle(1.5 and 0.75);
    \node[blobtext] at (1.5,-1.5) {$F$};
  \end{tikzfigure}
  \quad\,\Rightarrow\quad\,
  \begin{tikzfigure}{0.6}{0.5}
    \draw[gluon] (0.5,0) -- (0.5,-1.5);
    \draw[gluon] (1,0) -- (1,-1.5);
    \draw[gluon] (2.5,0) -- (2.5,-1.5);
    \node[gluontext] at (1.75,-0.45) {$\ldots$};
    \draw[wilson, implies-] (0,0) -- (3,0);
    \filldraw[wilsontext] (3,0) circle(0.125);
    \filldraw[blob] (1.5,-1.5) circle(1.5 and 0.75);
    \node[blobtext] at (1.5,-1.5) {$F$};
  \end{tikzfigure}
  \!= \; (-)^n \, \bar{C}
  \begin{tikzfigure}{0.6}{0.5}
    \draw[photon] (0.5,0) -- (0.5,-1.5);
    \draw[photon] (1,0) -- (1,-1.5);
    \draw[photon] (2.5,0) -- (2.5,-1.5);
    \node[photontext] at (1.75,-0.45) {$\ldots$};
    \draw[wilson, -implies] (0,0) -- (3,0);
    \filldraw[wilsontext] (0,0) circle(0.125);
    \filldraw[blob] (1.5,-1.5) circle(1.5 and 0.75);
    \node[blobtext] at (1.5,-1.5) {$F$};
  \end{tikzfigure}
  \ep
\end{equation}
The wavy lines maybe resemble photon lines, but are just a reminder that there is no color structure left in the blob.
Of course, a lot of blob structures won't be color factorable, but we can always write these as a sum of factorable terms:
\begin{equation*}
  F_{\mu_1\cdots\mu_n}^{a_1\cdots a_n} (k_1,\ldots, k_n) =
    \sum\limits_i C_i^{a_1\cdots a_n} F_{i\, \mu_1\cdots\mu_n} (k_1,\ldots, k_n)\ec
\end{equation*}
such that we can repeat the same procedure as before.

\section{Piecewise Linear Wilson Lines}\label{sec: PiecewiseLinear}
When connecting an $n$-gluon blob to a piecewise Wilson line, the $n$ gluons aren't necessarily all connected to the same segment, but can be divided among several segments.
When connecting e.g.\@  a $4$-gluon blob, we need to calculate exactly 5 diagrams (independent on the number of segments $M$). These diagrams are:
\begin{equation*}
  \mathcal{U}^J_4, \, \mathcal{U}^J_3\mathcal{U}^K_1, \, \mathcal{U}^J_2\mathcal{U}^K_2, \,
  \mathcal{U}^J_2\mathcal{U}^K_1\mathcal{U}^L_1, \, \text{and }
  \mathcal{U}^J_1\mathcal{U}^K_1\mathcal{U}^L_1\mathcal{U}^O_1.
\end{equation*}
They are the easiest represented schematically:
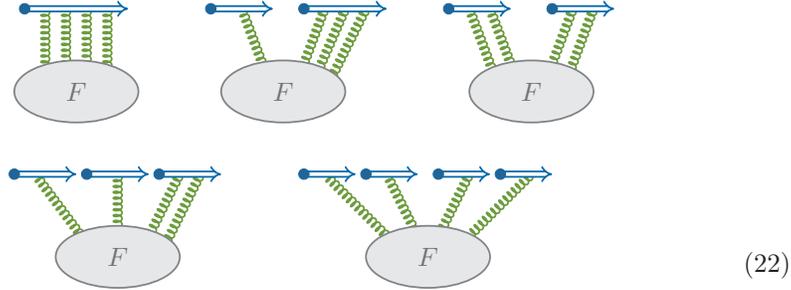
\begin{equation}\label{eq: 4gluondiags}
  \begin{tikzfigure}{0.55}{3.5}
    \draw[gluon] (0.75,0) -- (0.75,-2);
    \draw[gluon] (1.25,0) -- (1.25,-2);
    \draw[gluon] (1.75,0) -- (1.75,-2);
    \draw[gluon] (2.25,0) -- (2.25,-2);
    \draw[wilson, -implies] (0.25,0) -- (2.75,0);
    \filldraw[wilsontext] (0.25,0) circle(0.125);
    \filldraw[blob] (1.5,-2) circle(1.5 and 0.75);
    \node[blobtext] at (1.5,-2) {$F$};
    \begin{scope}[shift={(5,0)}]
      \draw[gluon] (0.5,0) -- (1.25,-2);
      \draw[gluon] (2.5,0) -- (1.75,-2);
      \draw[gluon] (3,0) -- (2.125,-2);
      \draw[gluon] (3.5,0) -- (2.5,-2);
      \draw[wilson, -implies] (-0.25,0) -- (1.25,0);
      \draw[wilson, -implies] (2,0) -- (4,0);
      \filldraw[wilsontext] (-0.25,0) circle(0.125);
      \filldraw[wilsontext] (2,0) circle(0.125);
      \filldraw[blob] (1.5,-2) circle(1.5 and 0.75);
      \node[blobtext] at (1.5,-2) {$F$};
    \end{scope}
    \begin{scope}[shift={(11,0)}]
      \draw[gluon] (0,0) -- (0.75,-2);
      \draw[gluon] (0.5,0) -- (1.25,-2);
      \draw[gluon] (2.5,0) -- (1.75,-2);
      \draw[gluon] (3,0) -- (2.25,-2);
      \draw[wilson, -implies] (-0.5,0) -- (1,0);
      \draw[wilson, -implies] (2,0) -- (3.5,0);
      \filldraw[wilsontext] (-0.5,0) circle(0.125);
      \filldraw[wilsontext] (2,0) circle(0.125);
      \filldraw[blob] (1.5,-2) circle(1.5 and 0.75);
      \node[blobtext] at (1.5,-2) {$F$};
    \end{scope}
    \begin{scope}[shift={(1,-4)}]
      \draw[gluon] (1.5,0) -- (1.5,-2);
      \draw[gluon] (3,0) -- (2,-2);
      \draw[gluon] (3.5,0) -- (2.5,-2);
      \draw[gluon] (-0.5,0) -- (1,-2);
      \draw[wilson, -implies] (0.75,0) -- (2.25,0);
      \draw[wilson, -implies] (-1,0) -- (0.5,0);
      \draw[wilson, -implies] (2.5,0) -- (4,0);
      \filldraw[wilsontext] (-1,0) circle(0.125);
      \filldraw[wilsontext] (2.5,0) circle(0.125);
      \filldraw[wilsontext] (0.75,0) circle(0.125);
      \filldraw[blob] (1.5,-2) circle(1.5 and 0.75);
      \node[blobtext] at (1.5,-2) {$F$};
    \end{scope}
    \begin{scope}[shift={(8.5,-4)}]
      \draw[gluon] (0.5,0) -- (1.5,-2);
      \draw[gluon] (-1,0) -- (1,-2);
      \draw[gluon] (2.5,0) -- (1.5,-2);
      \draw[gluon] (4,0) -- (2,-2);
      \draw[wilson, -implies] (0,0) -- (1.25,0);
      \draw[wilson, -implies] (-1.5,0) -- (-0.25,0);
      \draw[wilson, -implies] (1.75,0) -- (3,0);
      \draw[wilson, -implies] (3.25,0) -- (4.5,0);
      \filldraw[wilsontext] (0,0) circle(0.125);
      \filldraw[wilsontext] (-1.5,0) circle(0.125);
      \filldraw[wilsontext] (1.75,0) circle(0.125);
      \filldraw[wilsontext] (3.25,0) circle(0.125);
      \filldraw[blob] (1.5,-2) circle(1.5 and 0.75);
      \node[blobtext] at (1.5,-2) {$F$};
    \end{scope}
  \end{tikzfigure}
\end{equation}
To calculate diagrams with different path structures (as defined in \eqref{eq: basicstructures}), we can use the same trick as in the end of the former section, viz.\@ a sign change and an interchange of the corresponding color indices. For instance:
\begin{equation*}
  \begin{tikzfigure}{0.6}{0.75}
    \draw[gluon] (0,0) -- (0.75,-2);
    \draw[gluon] (0.5,0) -- (1.25,-2);
    \draw[gluon] (2.5,0) -- (1.75,-2);
    \draw[gluon] (3,0) -- (2.25,-2);
    \draw[wilson, -implies] (-0.5,0) -- (1,0);
    \draw[wilson, implies-] (2,0) -- (3.5,0);
    \filldraw[wilsontext] (-0.5,0) circle(0.125);
    \filldraw[wilsontext] (3.5,0) circle(0.125);
    \filldraw[blob] (1.5,-2) circle(1.5 and 0.75);
    \node[blobtext] at (1.5,-2) {$F$};
  \end{tikzfigure}
  \quad = \quad
  (-)^2
  \begin{tikzfigure}{0.6}{0.75}
    \draw[gluon] (0,0) -- (0.75,-2);
    \draw[gluon] (0.5,0) -- (1.25,-2);
    \draw[gluon] (2.5,0) -- (1.75,-2);
    \draw[gluon] (3,0) -- (2.25,-2);
    \draw[wilson, -implies] (-0.5,0) -- (1,0);
    \draw[wilson, -implies] (2,0) -- (3.5,0);
    \filldraw[wilsontext] (-0.5,0) circle(0.125);
    \filldraw[wilsontext] (2,0) circle(0.125);
    \filldraw[blob] (1.5,-2) circle(1.5 and 0.75);
    \node[blobtext] at (1.5,-2) {$ F \big|_{a_3\leftrightarrow a_4}$};
  \end{tikzfigure}.
\end{equation*}
The easiest way to implement this, is to define a path function $\Phi$ per diagram for a given blob, that gives the color structure in function of the path type. For the leading order 2 gluon blob, this is straightforward:
\begin{subalign}[eq: 2gluonpathconstants]
  \begin{tikzfigure}{0.6}{0.2}
    \draw[gluon] (0.25,0) to[out=-60,in=-120,distance=25] (2,0);
    \draw[wilson,-implies] (-0.25,0) -- (2.25,0);
    \filldraw[wilsontext] (-0.25,0) circle(0.125);
  \end{tikzfigure}
  &:
  \qquad \Phi(J) = C_F,
  \\
  \begin{tikzfigure}{0.6}{0.2}
    \draw[gluon] (0.5,0) to[out=-90,in=-90,distance=30] (2.5,0);
    \draw[wilson,-implies] (-0.5,0) -- (1,0);
    \filldraw[wilsontext] (-0.5,0) circle(0.125);
    \draw[wilson,-implies] (1.5,0) -- (3,0);
    \filldraw[wilsontext] (1.5,0) circle(0.125);
  \end{tikzfigure}
  &:
  \qquad \Phi(J,K) = (-)^{\phi_J + \phi_K}C_F,
\end{subalign}
where $\phi_J$ represents the structure of the segment:
\begin{equation}
  \phi_J = \begin{cases}
    0 & J = \wilsonup\\
    1 & J = \wilsondown
  \end{cases}\ep
\end{equation}
Keep in mind that in our original definition of the Wilson line \eqref{eq: Wilsondef}, color indices are not yet traced, hence equations \eqref{eq: 2gluonpathconstants} should still be multiplied with a unit matrix.
For non-factorable blobs we use the same trick as before, by giving $\Phi$ an extra index to identify the sub diagram it belongs to.

Let us introduce a new notation, to indicate a full diagram but without the color content, in which a blob is connected to $m$ Wilson line segments with $n_i$ gluons connected to the $i$-th segment:
\begin{equation}
  \mathcal{W}_{n_m\cdots n_1}^{J_m\cdots J_1}\ep
\end{equation}
We will write the indices from right to left for convenience.
Returning to the 4-gluon blob, we can now write the full result for a factorable blob, carefully summing the sub-diagrams to keep path-ordering\cite{VanderVeken:2014kna}:
\begin{multline}\label{eq: 4gluonblobresult}
  \mathcal{U}^4 = 
    \sum_J^M \Phi_4\mathcal{W}_{4}^{J} +
    \sum_{J=2}^M \sum_{K=1}^{J-1}
      \left[ \Phi_{3\,1} \mathcal{W}_{3\,1}^{JK} + \Phi_{2\,2} \mathcal{W}_{2\,2}^{JK} \right]
    +
    \sum_{J=3}^M \sum_{K=2}^{J-1} \sum_{L=1}^{K-1} \Phi_{2\,1\,1} \mathcal{W}_{2\,1\,1}^{JKL}\\
    +
    \sum_{J=4}^M \sum_{K=3}^{J-1} \sum_{L=2}^{K-1} \sum_{O=1}^{L-1} 
      \Phi_{1\,1\,1\,1} \mathcal{W}_{1\,1\,1\,1}^{JKLO} +
    \text{symm.}
\end{multline}
The diagrams $\Phi_{1\,3} \mathcal{W}_{1\,3}^{JK}$, $\Phi_{1\,2\, 1} \mathcal{W}_{1\,2\,1}^{JKL}$, and $\Phi_{1\,1\,2} \mathcal{W}_{1\,1\,2}^{JKL}$ which occur after symmetrization, are calculated using plain substitution, interchanging also the $\phi_J$.
For a non-factorable blob, every term is just replaced by a sum over sub diagrams.
It is important to realise that both the $\Phi_{n_i\cdots}$ and $\mathcal{W}_{n_i \cdots}$ can be calculated independently of the path structure, giving a result depending on $n_J$, $r_J$ and $\phi_J$, which can easily be ported to different path structures.

So far we have only calculated amplitudes of Wilson line diagrams. In QCD calculations, it is common to calculate probabilities directly, by using cut diagrams. This framework can easily be adapted to allow for this kind of calculations, splitting the sum in \eqref{eq: 4gluonblobresult} into three distinct sectors: a sector where the blob is only connecting segments left of the cut, a sector where the blob is only connecting segments right of the cut, and a sector where the blob is connecting segments both left and right of the cut. In other words:
\begin{equation}
  \mathcal{U} = \mathcal{U}_\text{left} + \mathcal{U}_\text{cut} + \mathcal{U}_\text{right}.
\end{equation}
For the left and right sectors nothing changes, the calculations go as before. For the example of the 4-gluon blob, the first sector $\mathcal{U}^4_\text{left}$ is almost exactly equal to \eqref{eq: 4gluonblobresult}, but the sums run up only to $M_c$, the number of segments before the cut, instead of $M$. The last sector $\mathcal{U}^4_\text{right}$ is simply the hermitian conjugate of this, starting at $M_c\!+\!1$.
For the remaining sector $\mathcal{W}_\text{cut}$ we expand our set of basic diagrams, adding diagrams where the Wilson line is connected to a cut blob. Several possible cut blobs might exist for one blob, depending on the number of gluons to the left and right of the cut. E.g.\@ the leading order 4-gluon cut blobs are given by
\begin{subalign}
  \begin{tikzfigure}{0.5}{1} \label{eq: 4gluoncutblob13}
    \draw[gluon] (0.3,-1) -- (1,-2.2);
    \draw[gluon] (1.7,-1) -- (1.2,-2.2);
    \draw[gluon] (2.2,-1) -- (1.7,-2.2);
    \draw[gluon] (2.7,-1) -- (2,-2.2);
    \filldraw[blob] (1.5,-2) circle(1. and 0.5);
    \draw[finalstatecut] (1.2,-2.75) -- (1.2,-0.9);
  \end{tikzfigure}
  &=
  \begin{tikzfigure}{0.5}{0.25}
    \draw[gluon] (0,0) to[out=-60, in=-120,distance=25] (1.5,0);
    \draw[gluon] (2,0) to[out=-60, in=-120,distance=25] (3.5,0);
    \draw[finalstatecut] (0.75,0.2) -- (0.75,-1);
  \end{tikzfigure}
  +\text{cross.},
  \\
  \begin{tikzfigure}{0.5}{1} \label{eq: 4gluoncutblob22}
    \draw[gluon] (0.3,-1) -- (1,-2.2);
    \draw[gluon] (1,-1) -- (1.2,-2.2);
    \draw[gluon] (2,-1) -- (1.7,-2.2);
    \draw[gluon] (2.7,-1) -- (2,-2.2);
    \filldraw[blob] (1.5,-2) circle(1. and 0.5);
    \draw[finalstatecut] (1.5,-2.75) -- (1.5,-0.9);
  \end{tikzfigure}
  &=
  \begin{tikzfigure}{0.5}{0.25}
    \draw[gluon] (0,0) to[out=-60, in=-120,distance=25] (1.5,0);
    \draw[gluon] (2,0) to[out=-60, in=-120,distance=25] (3.5,0);
    \draw[finalstatecut] (1.75,0.25) -- (1.75,-1);
  \end{tikzfigure}
  +
  \begin{tikzfigure}{0.5}{0.4}
    \draw[gluon] (0,0) to[out=-60, in=-120,distance=42] (2.5,0);
    \draw[gluon] (0.5,0) to[out=-60, in=-120,distance=25] (2,0);
    \draw[finalstatecut] (1.25,0.) -- (1.25,-1.5);
  \end{tikzfigure}
  +\text{cross.},
\end{subalign}
where the crossings are to be made on the sides of the cut separately. When the cut blob is more complex, it should be summed over all possible cut locations.

\section{Example Calculation}
Let us recapitulate our framework with a small example, viz.\@ the LO 2-gluon blob. At any order, there are 3 possible 2-gluon diagrams:
\begin{equation}
\begin{tikzfigure}{0.45}{0.8}
  \draw[gluon] (0.75,0) -- (0.75,-1.75);
  \draw[gluon] (2.25,0) -- (2.25,-1.75);
  \filldraw[blob] (1.5,-1.75) circle(1.5 and 0.75);
  \draw[wilson,-implies] (0,0) -- (3,0);
  \filldraw[wilsontext] (0,0) circle(0.125);
  \begin{scope}[shift={(5,0)}]
    \draw[gluon] (0.625,0) -- (1,-1.75);
    \draw[gluon] (2.325,0) -- (2,-1.75);
    \filldraw[blob] (1.5,-1.75) circle(1.5 and 0.75);
    \draw[wilson,-implies] (0,0) -- (1.25,0);
    \filldraw[wilsontext] (0,0) circle(0.125);
    \draw[wilson,-implies] (1.75,0) -- (3,0);
    \filldraw[wilsontext] (1.75,0) circle(0.125);
  \end{scope}
  \begin{scope}[shift={(10,0)}]
    \draw[gluon] (0.625,0) -- (1,-1.75);
    \draw[gluon] (2.325,0) -- (2,-1.75);
    \filldraw[blob] (1.5,-1.75) circle(1.5 and 0.75);
    \draw[wilson,-implies] (0,0) -- (1.25,0);
    \filldraw[wilsontext] (0,0) circle(0.125);
    \draw[wilson,-implies] (1.75,0) -- (3,0);
    \filldraw[wilsontext] (1.75,0) circle(0.125);
    \draw[finalstatecut] (1.25,-2.75) -- (1.25,0);
  \end{scope},
\end{tikzfigure}
\end{equation}
At NLO the blob is just an LO gluon propagator (here given in Feynman gauge):
\begin{equation}
  F = -\i (2\pi)^\omega \diracdelta{k_1+k_2} \delta^{ab} \frac{1}{k_1^2+\i\varepsilon}g^{\mu_1\mu_2},
\end{equation}
and the cut blob is just a radiated gluon that is integrated over:
\begin{equation}
  F = - (2\pi)^{\omega+1} \diracdelta{k_1+k_2}\delta^{ab} \heavisidetheta{k_1^+}\diracdelta{k_1^2} g^{\mu\nu}.
\end{equation}
The color factors are given by \eqref{eq: 2gluonpathconstants}. We show here the result for the second diagram at $r_J=r_K$. If both segments are on-LC, the result is:
\begin{equation}
	\mathcal{W}_{1\,1}^{JK} \big|_{\text{LC}} =
		\frac{\alpha_s}{2\pi}
		\left( \frac{1}{\epsilon^2} + \frac{\pi^2}{3} \right)
		\left(
			\frac{n_K\!\cdot\! n_J}{2}\frac{\mu^2}{\eta^2}
		\right)^{\!\epsilon}
		\ep
\end{equation}
If both segments are off-LC, the result can be expressed in function of the angle $\chi$:
\begin{equation}
	\cosh \chi = \frac{n_K\cdot n_J}{|n_K|\,|n_J|}\ep
\end{equation}
For simplicity we normalize the directions, i.e.\@ $|n_K|=|n_J|$. This gives:
\begin{subalign}
\mathcal{W}_{1\,1}^{JK} \big|_{\text{\sout{LC}}} &=
	\frac{\alpha_s}{2\pi}\chi\text{coth}\chi\,
	\left(\frac{1}{\epsilon}+\Upsilon\right)
	\left[
		\frac{1}{4} n_K^2n_J^2 \sinh^2\!\chi\,
		\frac{\mu^2}{\eta^2}
	\right]^{\!\epsilon} \ec\\
\Upsilon &= 2\ln 2 \!+\! \ln n_K^2\!+\!2\ln(1\!+\!\exp{\chi})\!+\!\chi
		\!-\!\frac{1}{\chi}\left(
			\text{Li}_2\,\exp{\chi}\!-\!\text{Li}_2\,\exp{-\chi}
		\right) \ep
\end{subalign}
If $|n_K|\neq|n_J|$, only $\Upsilon$ is affected and will be a bit more involving.

\section*{Acknowledgements}
I am very grateful to the organisers of the QCD Evolution 2014 conference for creating such a fruitful environment. Furthermore I would like to thank I.O.~Cherednikov, M.~Echevarria, L.~Gamberg, A.~Idilbi, T.~Mertens, A.~Prokudin and P.~Taels for useful discussions and insights.

\bibliographystyle{ws-ijmpcs}
\bibliography{Bibliography}

\end{document}